\documentclass[11pt]{article}
\usepackage{graphicx}
\usepackage{amsmath}
\usepackage{hhline}
\usepackage{amssymb}
\usepackage{times}
\usepackage{cite}
\usepackage{array}
\usepackage{multirow}

\usepackage{color}
\definecolor{rltred}{rgb}{0.75,0,0}
\definecolor{rltgreen}{rgb}{0,0.5,0}
\definecolor{rltblue}{rgb}{0,0,0.75}

\newlength{\dinwidth}
\newlength{\dinmargin}
\begin{document}
\begin{titlepage}

\noindent
Date:        July 29, 2008       \\
                
\vspace{2cm}

\begin{center}
\begin{Large}

{\bf Proton Driven Plasma Wakefield Acceleration}

\vspace{2cm}

A.~Caldwell$^1$, K.~Lotov$^2$, A.~Pukhov$^3$, F.~Simon$^{1,4}$ \\

\end{Large}

\vspace{2cm}
$^1$Max-Planck-Institut f\"ur Physik, 80805 M\"unchen, Germany \\
$^2$Budker Institute of Nuclear Physics, 630090 Novosibirsk, Russia, \\
Novosibirsk State University, 630090 Novosibirsk, Russia\\
$^3$Institut f\"ur Theoretische Physik I, \\
Heinrich-Heine-Universit\"at D\"usseldorf, 40225 D\"usseldorf, Germany \\
$^4$Excellence Cluster `Origin and Structure of the Universe', Garching, Germany

\end{center}

\vspace{2cm}

\begin{abstract}
Plasma wakefield acceleration, either laser driven or electron-bunch driven, has been demonstrated to hold great potential.  However, it is not obvious how to scale these approaches to bring particles up to the TeV regime. In this paper, we discuss the possibility of proton-bunch driven plasma wakefield acceleration, and show that high energy electron beams could potentially be produced in a single accelerating stage.
\end{abstract}
\end{titlepage}

\section{Introduction}
It has been known for some time that plasmas can support very large electric fields, and can therefore be used for accelerating particles to relativistic energies.  Initially, laser driven plasma wakefield acceleration was considered in the literature~\cite{Taj}, and experimental verification of the ideas followed~\cite{PRL47-1285,PRL68-48,AIP335-145}.  Detailed simulations of the process are now available which have indicated  the production of electron beams with interesting characteristics.  In recent experiments, gradients in the range $10-100$~GV/m have been achieved. These have so far been limited to distances of a few cm, but the progress has been very impressive~\cite{ref:GeV}.  In order to accelerate an electron bunch to 1~TeV, these gradients would have to be maintained over distances of tens of meters, or many acceleration stages would have to be combined. 

It was later recognized that the plasma could also be excited by an electron bunch~\cite{Chen}.  Given an intense enough bunch of electrons, the plasma is both created~\cite{PRST-AB9-101301} and excited by the passage of the bunch.  Very large electric fields were predicted and later observed~\cite{PRL93-014802}.  In the linear regime, the maximal achievable gradient can be written~\cite{PoP9-1845} as
\begin{equation}
\label{eq:LWFA}
E=240 ({\rm MV/m}) \left(\frac{N}{4\cdot 10^{10}}\right)\left( \frac{0.6}{\sigma_z ({\rm mm})}\right)^2
\end{equation}
where $N$ is the number of particles in the driving bunch and $\sigma_z$ is the length (rms) of the bunch assuming a Gaussian beam profile. In the case of electron driven plasma wakefield acceleration, a gradient of $50$~GV/m was achieved and sustained at SLAC for almost $1$~m~\cite{ref:Blumenfeld}.  However, the maximum energy which can be given to a particle in the witness bunch is limited by the transformer ratio, 
$$R=\frac{E_{\rm max}^{\rm witness}}{E_{\rm max}^{\rm drive}}\leq 2-\frac{N_{\rm witness}}{N_{\rm drive}}$$
which is at most $2$ for longitudinally symmetric drive bunches~\cite{ref:Ruthetal}.  This upper limit can in principle be overcome by nonsymmetric bunches~\cite{ref:asymdrive}, but, given the difficulty in producing a short bunch in the first place, it is hard to imagine that these short bunches can also be shaped and the shape maintained over the distances needed for acceleration to high energies. $500$~GeV electron beams would therefore be needed to produce a $1$~TeV beam, or many stages of plasma acceleration would be needed to arrive at the desired energy range.  In the first case, one could imagine a bootstrapping process, where a drive bunch of $50$~GeV is used to produce $100$~GeV bunches, which are then used to produce $200$~GeV bunches, etc.  The efficiency of such a process would be an issue.  In the second case, very precise control of the bunch exiting the plasma would be required since the relative timing of the drive bunch and witness bunch is critical. 

Plasma wave excitation by a negatively charged driver is now well studied both theoretically \cite{ref:Esarey,ref:Joshi,ref:Katsouleas,nonlin} and experimentally \cite{ref:Esarey,ref:Blumenfeld,ref:Muggli}.  In contrast to plasmas driven by electron beams, only limited investigations of  the plasma wave excitation by a positively charged driver exist~\cite{PRE64-045501,LPB21-497,PoP13-092109,PRL90-205002,PRL90-214801}. In the linear wake field regime the electric field distribution should be the same as that for the negative driver but shifted in phase.  However, the path to the non-linear stage differs significantly as revealed in multi-dimensional particle-in-cell (PIC) simulations~\cite{ref:WeiLu}. Physically, the negatively charged driver ``blows out'' the background plasma electrons creating a low density region behind the driver. The non-linear ``blow out'' regime has quite useful properties that make it successful in plasma-based acceleration: it provides the very high accelerating field, which does not depend on the transverse coordinate, while the transverse fields are focusing both for the driver and for the witness bunch. The ``blow-out'' regime allows for a rapid energy transfer from the driver to the witness beam, thus leading to an efficient acceleration of electrons~\cite{ref:lotov_energy_efficient}. 

It is much more difficult to reach the blow-out regime with a positively charged driver, such as protons. Instead of ``blowing out'' plasma electrons, they ``suck them in'' towards the propagation axis. Due to the radial symmetry, this leads to an electron density enhancement on-axis and effective increase of the local plasma frequency. As a result, the proton driver must be even shorter in order to excite the plasma wake resonantly. In the highly non-linear regime, one has to rely on numerical simulations to find the optimal beam-plasma parameters for the resonant wake field generation.

In this paper, we investigate the possibility of driving the plasma wave with an intense bunch of protons.  Given that protons can be accelerated to the TeV regime in conventional accelerators, it is conceivable to accelerate electron bunches in the wake of the proton bunch up to several TeV (e.g., in the wake of an LHC proton beam) in one pass through the plasma.  Several issues come to mind, such as the possibility of producing a proton bunch with large enough charge density, the possibility of phase slippage as protons slow down, the effect of proton beam divergence and dissipation in the plasma, etc.  We will discuss these points below.  We then discuss a specific parameter set close to existing proton beams, and show that the production of a $O(1)$~TeV electron beam is in principle allowed with a PDPWA (proton driven plasma wakefield accelerator). 

\section{Initial Considerations}

\begin{figure}[hbpt]
\begin{center}
\includegraphics[width=0.99\textwidth]{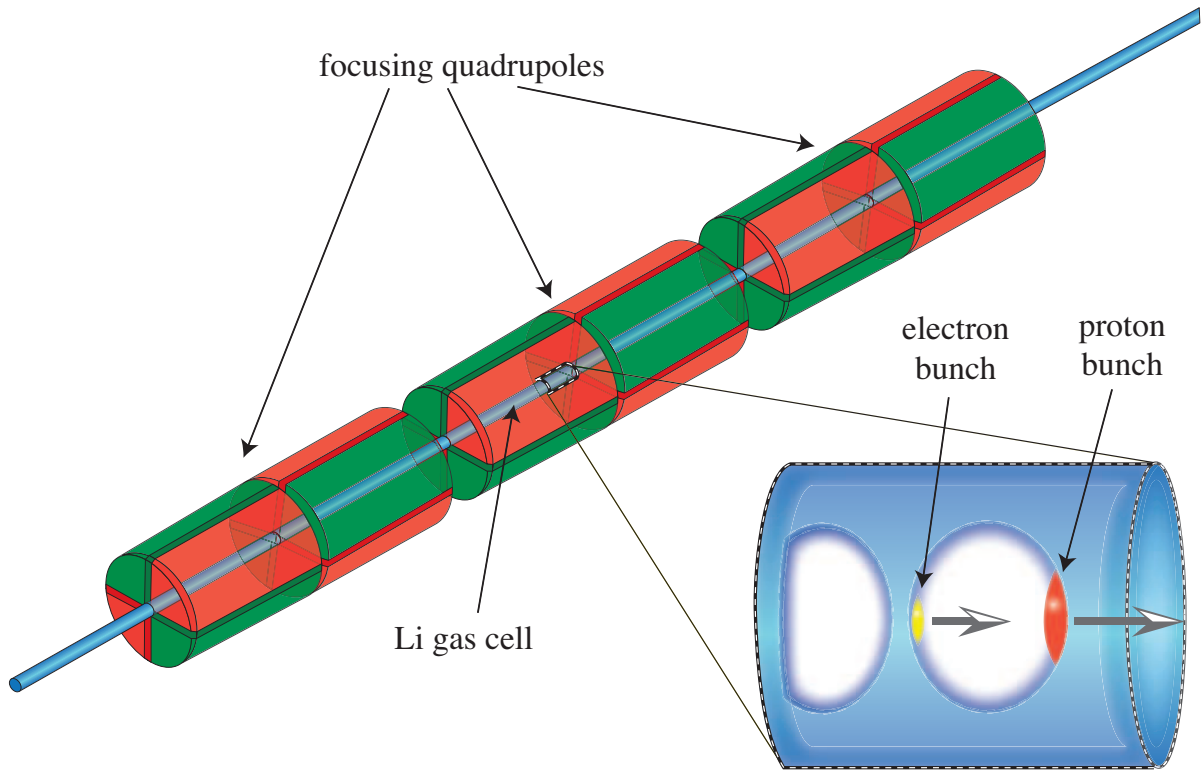}
\caption{\it A schematic description of a  section of the plasma wakefield accelerating structure.  A thin tube ($5$~mm diameter)  containing Li gas is surrounded by quadrupole magnets with alternating polarity. The quadrupole magnets are $70$~cm long.  The blow-up shows the plasma bubble created by the proton bunch (red). The electron bunch (yellow) undergoing acceleration is located at the back of the bubble. Note that the dimensions are not to scale.}
\label{fig:cell} 
\end{center}
\end{figure}

The accelerating structure we have studied for proton-driven plasma wakefield acceleration is given in Fig.~\ref{fig:cell}.  A high density proton bunch propagates through the plasma and sets the plasma electrons in motion. The plasma of the required density can be produced, e.g., by tunnel ionization of a neutral gas with a high-intensity laser pulse that propagates shortly ahead of the proton beam.  For a highly relativistic driving bunch, the electric field seen by the plasma electrons is in the transverse direction, and the plasma electrons begin to oscillate around their equilibrium position with frequency $\omega_p$ given by
$$\frac{\omega_p}{c} = \sqrt{\frac{n_p e^2}{\epsilon_0 mc^2}}$$
where $n_p$ is the density of plasma electrons.  Given their large mass, the plasma ions are effectively frozen.  The oscillating electrons initially move toward the beam axis, then pass through each other, creating a cavity with very strong electric fields. The cavity structure repeats, and the pattern moves with the proton bunch velocity.  An appropriately timed witness bunch can be placed in a region of very strong electric field and accelerated.  The plasma has focusing properties for the tail of the drive bunch, as well as for the witness bunch.

In the linear regime, Eq.~\ref{eq:LWFA} applies to particles of either charge, and can therefore be used to calculate the field produced by a bunch of protons passing through a cold plasma (the nonlinear case will be discussed below).  Proton bunches with $10^{11}$ protons per bunch are available today, and the main issue in producing strong electric fields in the plasma is the formation of short proton bunches.  TeV proton beams typically have a relative momentum spread $\sigma_p/p=10^{-4}$ and a rms bunch length of $50$~cm.  Assuming this longitudinal phase space area is preserved and that a technically feasible scheme for a phase rotation will be found, a proton bunch with $\sigma_z=100$~$\mu$m would have a momentum spread of about $50$~\%.  This is likely too large.  The LHC foresees reaching $\sigma_p/p=10^{-4}$ for bunches with rms length of only $7.55$~cm.  In this case, $\sigma_z=100$~$\mu$m could in principle be achieved with a momentum spread of about $7.5$~\%, which is much more favorable.  
We have simulated a $1$~TeV proton driver with 
$\sigma_z=100$~$\mu$m and $\sigma_p/p=0.1$.

\subsection{Longitudinal Growth of the Drive Bunch}
As the proton bunch propagates, the momentum spread will induce a longitudinal growth in the beam. This can be evaluated for vacuum propagation as follows:
$$d\approx \frac{L}{2\Delta \gamma^2} \approx \left(\frac{\sigma_p}{p}\right) \frac{M_P^2c^4}{p^2c^2}L \; ,$$
where $d$ is the spread of the bunch induced by the momentum spread, $L$ is the distance travelled, $M_P$ is the proton mass and $p$ the proton momentum. Given a $1$~TeV proton beam, $10$~\% momentum spread leads to a growth of about $0.1$~$\mu$m/m.  Large relative momentum spreads will still allow for long plasma acceleration stages provided the drive beam is relativistic.

\subsection{Transverse Growth of the Drive Bunch}
As mentioned above, the plasma has a strong focusing effect on the tail of the drive bunch, as well as on the witness bunch.  However, the head of the drive bunch will tend to fly apart unless quadrupole focusing is applied.  We therefore foresee an arrangement with strong focusing of the proton drive bunch along the length of the plasma channel.  A possibility for these quadrupoles are small diameter permanent magnets, such as those described in ~\cite{ref:Habs}, producing gradients of order 1~T/mm.

\subsection{Phase Slippage}
Another issue is phase slippage between the proton driving bunch and the electron witness bunch.  As the proton bunch travels through the plasma, it will slow down and the phase relation with the light electron bunch will begin to change.  The phase change is given by~\cite{ref:Ruthetal}
\begin{eqnarray*}
\delta & \approx & \frac{\pi L}{\lambda_p}\left[ \frac{1}{\gamma_f\gamma_i}\right] \\
           & \approx & \frac{\pi L}{\lambda_p} \left[ \frac{M_p^2c^4}{p_fp_ic^2}\right] \; ,
\end{eqnarray*}
where $\lambda_p$ is the plasma wavelength, and $p_{i,f}$ are the initial and final momenta of the protons in the driving bunch.  To maximize the gradient (Eq.~\ref{eq:LWFA}), the plasma wavelength should have a definite relation to the length of the driving bunch:
$$\lambda_p=\sqrt{2}\pi \sigma_z \;\; .$$
Requiring a phase slippage of only a fraction of the plasma wavelength implies that the driving beam energy cannot change appreciably,  and this could be a severe limitation on acceleration by a proton bunch.  However, with an initial proton energy of $1$~TeV and a final energy of $0.5$~TeV, it should still be possible to have plasma lengths of many meters.  In addition, it is possible to control the plasma wavelength by adjusting the density of the plasma~\cite{ref:Pukhov}, in part or fully compensating for the phase slippage.  

\subsection{Proton Interactions in the Plasma Cell}

Proton interactions in the plasma are not expected to be a big issue.  The plasma density and plasma wavelength are related by
\begin{eqnarray*}
\lambda_p & = & 2\pi\sqrt{\frac{\epsilon_0 mc^2}{n_pe^2}} \\
                    &\approx& 1~{\rm mm} \sqrt{\frac{10^{15} {\rm cm}^{-3}}{n_p}}  \; .
\end{eqnarray*}
 Typical values of $n_p$ will be in the range  $10^{14}-10^{17}$~cm$^{-3}$, and the mean-free-path for inelastic reactions of high energy protons with the gas will be orders of magnitude larger than the expected plasma cell length.  A Geant4~\cite{ref:Geant4} simulation for a $1$~TeV proton beam in Li vapor of density $1\cdot 10^{15}$ atoms/cm$^3$ gives a transverse growth rate of the proton beam of less than $0.01$~$\mu$m/m due to multiple scattering, which is small compared to the size of the proton bunch.

Assuming the same gas density, the amount of energy deposited in the magnets by a bunch of $10^{11}$ protons with $E_P=1$~TeV  due to beam secondaries is calculated to be only $150$~Gy/year, which should not pose a problem with activation of the magnet material or possible demagnetization in case permanent magnets are used.  Assuming the total proton bunch energy is uniformly distributed in a gas cell of length $400$~m, the protons would only deposit about $40$~J/m.  A string of $1000$ bunches may be passed through the cell in several seconds, so that up to $10$~kW of power could be deposited per meter over this time period.  This is not expected to pose any heating problems with the magnet system (assuming warm magnets are used), since rather large volumes of steel will be present.

The main issue in producing strong electric fields in the plasma and using them over long distances to accelerate electrons is clearly the need to produce a proton bunch of order $100$~$\mu$m in length or shorter.  

\section{Simulation of Plasma Wave Excitation by Relativistic Protons}

We have simulated the wakefield acceleration process assuming a pipe containing Li gas inserted in a region of alternating vertically focusing and defocusing quadrupole fields.    The parameters of the simulation are given in Table~\ref{tab:params}.  It was assumed that the gas was fully ionized. 
 
\begin{table}[htdp]
\caption{Table of parameters for the simulation.}
\begin{center}
\begin{tabular}{|r|c|c|c|} 
\hline
Parameter & Symbol & Value & Units \\
\hline
Protons in Drive Bunch & $N_P$ & $10^{11}$ & \\
Proton energy & $E_P$ & $1$ & TeV \\
Initial Proton momentum spread & $\sigma_p/p$ & $0.1$ & \\
Initial Proton longitudinal spread & $\sigma_Z$ & $100$ & $\mu$m \\
Initial Proton bunch angular spread & $\sigma_{\theta}$ & $0.03$ & mrad \\
Initial Proton bunch transverse size & $\sigma_{X,Y}$ & $0.4$ & mm \\
\hline
Electrons injected in witness bunch & $N_e$ & $1.5\cdot 10^{10}$ & \\
Energy of electrons in witness bunch & $E_e$ & $10$ & GeV\\
\hline
free electron density & $n_p$ & $6\cdot 10^{14}$ & cm$^{-3}$ \\
Plasma wavelength & $\lambda_p$ & $1.35$ & mm\\
Magnetic field gradient & & $1000$ & T/m \\
Magnet length & & $0.7$ & m \\
\hline
\end{tabular}
\end{center}
\label{tab:params}
\end{table}%

We have used the three-dimensional (3D) fully electromagnetic relativistic PIC code VLPL \cite{ref:vlpl} as well as the quasi-static PIC code LCODE~\cite{ref:lcode,lctxt} to simulate the beam-plasma interaction. The 3D PIC simulation gives a very detailed shape of the plasma wave and cross-checks the radially symmetric results of the quasi-static code. On the other hand, the computationally efficient quasi-static code allows to simulate electron acceleration up to TeV energies over hundreds of meters of plasma.

In our numerical simulations we have selected a bunch of $10^{11}$ protons with 1~TeV energy and transverse emittance $0.01$~mm$\cdot$mrad. The bunch has a Gaussian shape of length $\sigma_z=100\mu$m. According to the linear wake field formula, the optimal plasma density for this beam should be around $5\times 10^{15}$cm$^{-3}$. However, our simulations have revealed that due to the wake field nonlinearities, the proton beam excites a very poor plasma wave at this ``linearly predicted'' density. Scanning over a range of plasma densities, we found that the optimal wake is produced at the plasma density roughly ten times lower, $n_p=6\times 10^{14}$cm$^{-3}$. The corresponding plasma wavelength is $\lambda_p=2\pi c/\omega_p = 1.35$~mm.

The proton beam transverse size and angular spread also needed optimization. As expected, the plasma wave focuses and guides the tail of the bunch. However, the plasma field was too low to guide the head of the bunch.  Without additional
focusing, the proton bunch head would diffract over a distance of a few meters, whereas for TeV acceleration one needs hundreds meters of propagation.  To overcome the natural beam diffraction, we simulated a magnetic quadrupole guiding system as discussed above (see also Fig.~\ref{fig:cell}). The magnetic field gradient was taken to be 1~T/mm.  Given the quadrupole strength and the assumed proton
beam emittance, we can calculate the required beam radius of
$\sigma_r=0.43~$mm=$2~k_p^{-1}$, where $k_p=\omega_p/c$. This driver radius is not matched with the plasma density. Thus, the tail of the driver is subject to transverse betatron oscillations at the beginning of acceleration. Yet, after a few betatron periods it reaches dynamic equilibrium and becomes matched, while its head is guided by the quadrupoles.

\begin{figure}[htdp]
\includegraphics[width=0.99\textwidth]{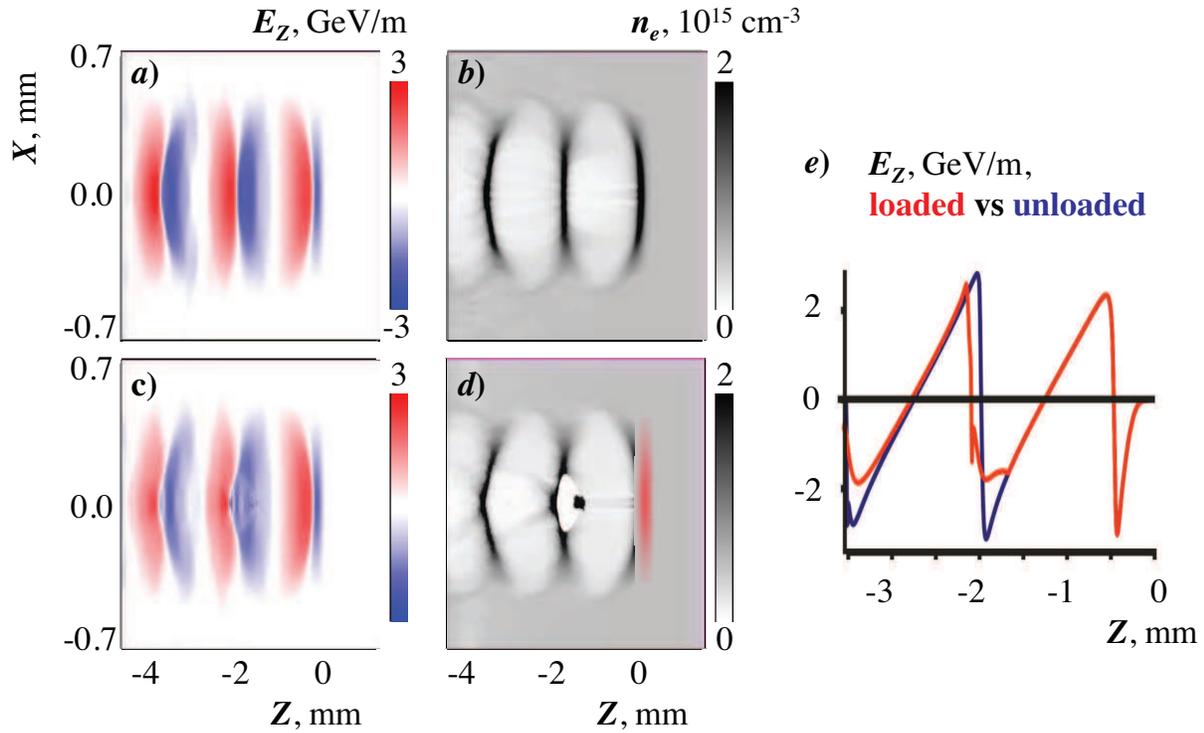}\caption{\label{fig:vlpl}
\it $(X-Z)$~cuts of the plasma wave electric field, frames (a), (c) and
electron density, frames (b), (d). Frames (a), (b) show unloaded
and frames (c), (d) loaded cases. The high density witness bunch is
seen as the black spot in the first wave bucket, frame (d). Frame (d) also shows the driving
proton bunch at the wave front (red).  Frame (e) gives the on-axis
accelerating field of the plasma wave for the unloaded (blue curve) and
loaded (red curve) cases. When the
wave is loaded, the accelerating field
magnitude decreases and the field flattens over the whole matched witness
beam.}
\end{figure}

The plasma wave generated by the proton driver is shown in Fig.~\ref{fig:vlpl}. The rightmost region of high electron density in frames b) and d) result from plasma electrons being ``sucked in'' by the proton bunch.  The electrons then continue to move across the beam axis and create a depletion region very similar to the blow-out region seen in the case of the electron driver. The electron witness bunch is placed on the left edge of the first bubble, where the longitudinal fields are strongest. The maximum accelerating field of the  wave is about $3$~GeV/m as shown in Fig.~\ref{fig:vlpl}e).  The accelerating gradient in this region increases as one approaches the left edge of the bubble, providing a stable regime of acceleration of the electron bunch.  The transverse electric fields in this region are also strong and act to focus the witness bunch.

To reach an energetically efficient regime of acceleration, we have to load the plasma wave with a matched witness bunch. Thus, the maximum accelerating field of the loaded wave will be somewhat lower. In our PIC simulations, we chose a witness electron bunch with $1.5\times 10^{10}$~particles and initial energy of $10$~GeV. The electric field and electron density from the loaded plasma wave are shown in Fig.~\ref{fig:vlpl}c)-d). The maximum accelerating field is about $1.7$~GeV/m and is nearly constant over the witness bunch.

\begin{figure}[htdp]
\includegraphics[width=0.99\textwidth]{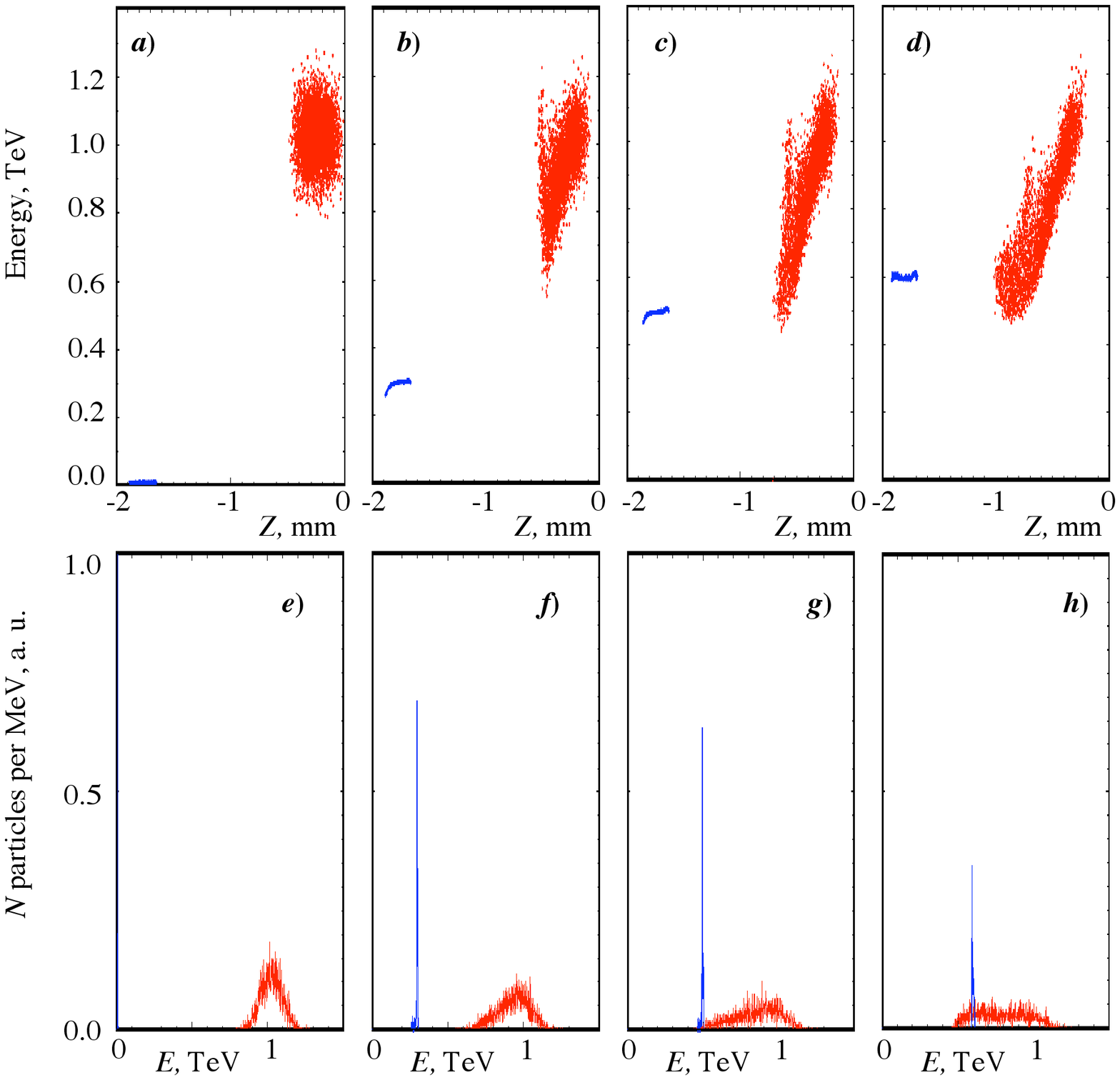}\caption{\label{fig:lcode_phasespace}
\it Snapshots of the combined longitudinal phase space of the driver and the witness beam
(energy vs coordinate), frames (a)-(d) and corresponding energy
spectra, frames (e)-(h). The snapshots are taken at acceleration
distances $Z=0,~150,~300,~450$~m.   The electrons are shown as blue points, while the protons are depicted as red points.
}
\end{figure} 

The acceleration over hundreds of meters of plasma has been simulated using the quasi-static code LCODE. Figure~\ref{fig:lcode_phasespace}a)-d) shows snapshots of the particle phase space (energy versus distance from the front of the proton bunch) at several distances along the channel. The proton bunch is initially distributed around $1$~TeV, while the electron bunch has a fixed energy of $10$~GeV.   Further down the channel, it is seen that the tail of the proton bunch loses significant amounts of energy, while the electron bunch picks up energy. Fig.~\ref{fig:lcode_phasespace}e)-h) show the energy spectra of the driver and of the witness bunches at different locations along the plasma channel.  

\begin{figure}[hbpt]
\includegraphics[width=0.99\textwidth]{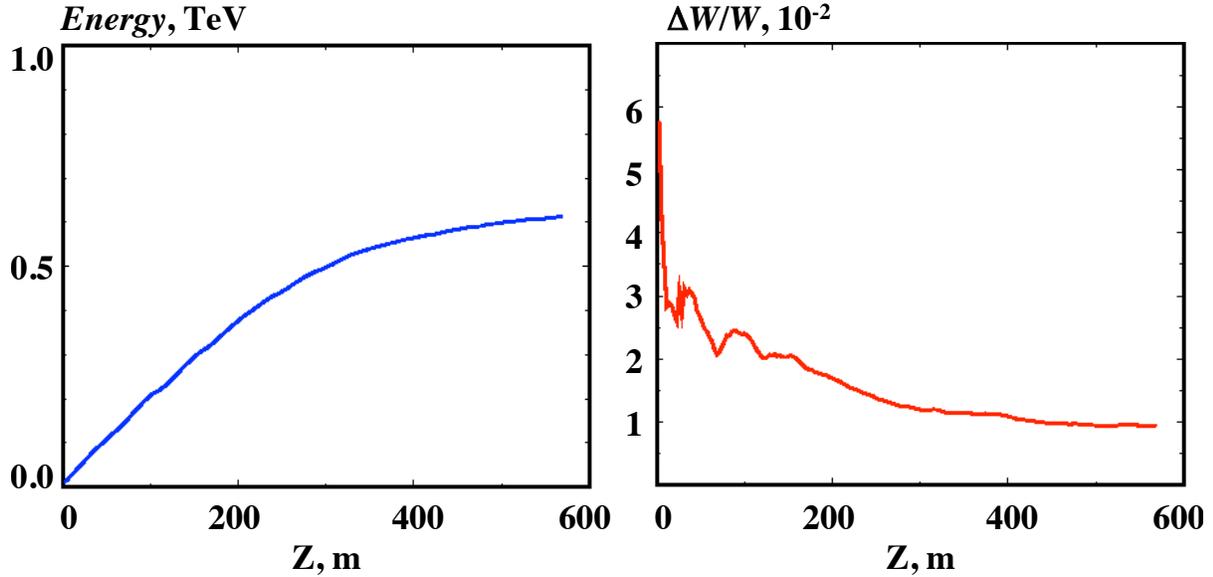}
\caption{\label{fig:energy}
\it Mean energy and energy spread of the witness beam as a
function of the acceleration distance.   }
\end{figure}

The mean energy of the electron bunch as a function of the distance along the channel is shown in Fig.~\ref{fig:energy}. After $450$~m of acceleration, the electron bunch reaches a mean energy of $0.62$~TeV per electron. The spread in the electron energy is also shown in Fig.~\ref{fig:energy}, and is about $1$~\% at the highest energies. This value could likely be improved with optimization of the witness bunch shape.  The overall energy conversion from the driver bunch to the witness bunch after this distance was nearly 10\%.  As can be clearly seen in Fig.~\ref{fig:lcode_phasespace}a)-d),  the proton bunch phase space changes considerably over the length of the channel, and the acceleration of the electron bunch decreases significantly after about $400$~m. The proton bunch acquires a large spread in both momentum and position. After 450~m propagation, the proton bunch length grows so much that it leaves the resonance condition and the plasma wave excitation becomes inefficient.  

The normalized transverse emittance of the electron bunch is not adversely affected by the plasma acceleration.  However, it should be noted that the scattering of electrons on plasma ions was not included in the PIC simulations.  A separate simulation using GEANT4 indicates that the growth in emittance from this effect will be tolerable.  Furthermore, no degradation of the emittance resulting from a resonance of the betatron oscillations of the witness bunch with the periodic focusing system was observed in the PIC simulations.  This may be due to the very rapid acceleration of the witness bunch, such that the emittance has no time to grow on passing through dangerous energy intervals.

\section{Discussion}

The simulation results indicate that a proton bunch could indeed be used to accelerate a bunch of electrons to high energies.  Further tuning of parameters would likely lead to improvements in simulation results. The key issue for the future applicability of proton-driven plasma wakefield acceleration will be the ability to phase rotate a high energy bunch of protons in such a way that the bunch is very short, of order $100$~$\mu$m or less.  Clearly, advances in longitudinal proton beam cooling would make this task much simpler.

 The acceleration of positrons has not been addressed here, and could be considerably more difficult than the acceleration of electrons~\cite{Lotov2007}. Initial investigations indicate that the electric field configurations do not have the broad equilibrium region seen for electron bunches, such that achieving a stable acceleration regime will be challenging.
 
 Achieving sufficient luminosities for an $e^+e^-$ collider would require a high repetition rate for producing high energy proton bunches, as well as strong focusing of the high energy electron and positron bunches.  Efficient energy transfer from the proton bunch to the electron bunch was observed in our simulations, and improvements, such as using plasma channels, are expected to further increase the efficiency.
 
 There are clearly many challenges to the development of a proton driven plasma wakefield accelerator.  Given the potential of proton driven plasma wakefield acceleration demonstrated in this paper, these challenges should be taken up and hopefully resolved.

\section{Acknowledgments}
We would like to thank S. Chattopadhyay, E. Elsen and F. Willeke  for useful discussions concerning proton bunch compression.

This work has been supported in part by the Russian Science Support Foundation, Russian President grants MD-4704.2007.2 and NSh-2749.2006.2, RFBR grant 06-02-16757, and the Russian Ministry of Education grant RNP.2.2.1.1.3653.

This work has also been supported by the DFG cluster of excellence `Origin and Structure of the Universe'.

\end{document}